\documentclass[twocolumn,preprintnumbers,aps]{revtex4}
\usepackage{dcolumn}
\begin{document}
\title{Lattice expansion and non-collinear to collinear ferrimagnetic order in MnCr$_2$O$_4$ nanoparticle}
\author{R.N. Bhowmik{\footnote{e-mail: rnb@cmp.saha.ernet.in}}}
\altaffiliation{Present address: Department of Physics, Mahishadal Raj College, Mahishadal, East Midnapore, Pin Code: 721628, West Bengal, India}
\author{R.Ranganathan{\footnote{e-mail: ranga@cmp.saha.ernet.in}}}
\affiliation{Experimental Condensed Matter Physics division, Saha Institute of Nuclear Physics, Kolkata-700064, India}
\author{R.Nagarajan}
\affiliation{Condensed Matter Physics division, Tata Institute of Fundamental Research, Colaba Road,Mumbai,India}

\begin{abstract}
We report magnetic behaviour of MnCr$_2$O$_4$, which belongs to a special class of spinel, 
known as chromite. Bulk MnCr$_2$O$_4$ shows a sequence of magnetic states, which follows paramagnetic 
(PM) to collinear ferrimagnetic (FM)
state below T$_C$ $\sim$ 45 K and collinear FM state to non-collinear FM state below T$_S$ $\sim$ 18 K. The non-collinear spin structure has been modified on decreasing the particle size, and 
magnetic transition at T$_S$ decreases in nanoparticle samples. 
However, ferrimagnetic order is still dominating in nanoparticles, except the observation of superparamagnetic like blocking and decrease of spontaneous magnetization for nanoparticle. This may, according to the core-shell model of ferrimagnetic nanoparticle, be the surface disorder effect of nanoparticle. The system also show the increase of T$_C$ in nanoparticle samples, 
which is not consistent with the core-shell model.   
The analysis of the M(T) data, applying spin wave theory,
has shown an unusual Bloch exponent value 3.35 for bulk MnCr$_2$O$_4$, which 
decreases and approaches to 1.5, a typical value for any standard ferromagnet, with 
decreasing the particle size. MnCr$_2$O$_4$ has shown a few more unusual behaviour. 
For example, lattice expansion in nanoparticle samples.
The present work demonstrates the correlation between a 
systematic increase of lattice parameter and the gradual decrease of B site 
non-collinear spin structure in the light of magnetism of MnCr$_2$O$_4$ 
nanoparticles. 
\end{abstract}
\maketitle

\section{Introduction}
Spinel ferrite \cite{Novak}, represented by formula unit AB$_2$X$_4$, has  generated renewed research interest
due to applications in nanoscience and technology. It has been established 
that the physical properties of spinel ferrites depend on the distribution of cations amongst 
the Tetrahedral (A) and Octahedral (B) sites and relative strengths of various kind 
of superexchange interactions via anions (X: O, S, Se ions). It is noted that most of the works 
are confined on spinel ferrites MFe$_2$O$_4$ (M= Zn,  Co, Mn, Ni etc.) \cite{Dorman}. In 
these spinels B sites are occupied by Fe$^{3+}$. There are many spinels of different 
class and having no B site Fe$^{3+}$, but may be very relevant in view of physics and 
technological application. 
Chromites MCr$_2$X$_4$ (M = Mn, Fe etc., X= O, S ions) are one of such classes which exhibits 
many unusual magnetic properties like ferrimagnetism, colossal magneto resistance (CMR) 
effect etc. Recently, extensive work on Half-metallic ferrimagnet (HFM) FeCr$_2$S$_4$ shows varieties 
of magnetic properties with different (magnetic and non-magnetic) 
substitutions \cite{Cava,Park}. It has been proposed \cite{Park} that Mn[Cr$_{2-x}$V$_x$]S$_4$ series 
might be a strong candidate for HFM. One would expect similar magnetic and transport 
properties in MCr$_2$O$_4$ in comparison with MCr$_2$S$_4$, since both (sulfide and oxide) 
spinels have identical cubic lattice structure, and both are ferrimagnet with T$_C$ in the range (60 K-80 K) and  non-collinear ferrimagnet
below 20 K. The other interests 
for investigating the MnCr$_2$O$_4$ spinel is to understand the role of strongly negative J$_{BB}$ (Cr-O-Cr) interactions in controlling the magnetic properties of chromites. In collinear ferrimagnetic structure of spinel A-O-B and B-O-B bond angle are 125$^0$ and 90$^0$, respectively.
For other configurations the distance between the oxygen ion and cations
are too large to give rise to a strong A-O-B superexchange interaction. 
The investigation of the magnetic properties of chromites \cite{Wick} revealed a number of unconventional features which could not be explained by Neel's theory for collinear (ferrimagnetic) spin structure.\\
It was proposed by Wickham and Goodenough \cite{Wick} that direct (antiferromagnetic) interactions between B site cations 
are responsible for the non-collinear ferrimagnetic structure in chromites like MnCr$_2$O$_4$. In fact, many phenomena in recent years have been explained in terms of B site direct cation-cation interaction, as in ZnCr$_2$O$_4$ \cite{Lee}. 
Neutron diffraction experiment confirmed the non-collinear spin structure between B site Cr$^{3+}$ moments in chromites \cite{Tsur}, including MnCr$_2$O$_4$ \cite{Hasti}. The neutron experiment shows that the decrease of magnetization in MnCr$_2$O$_4$ below 18 K, though identical with re-entrant magnetic transition \cite{Sanjay}, originates from the  occuring of non-collinear spin structure in MnCr$_2$O$_4$.\\
Numerous attempts have been made to bridge the gap between ferrites (MFe$_2$O$_4$), which are collinear ferrimagnet, and chromites, which are non-collinear ferrimagnet. However, the picture of spin configuration of chromites are not very clear even upto date.
Therefore, an investigation for chromites is essential and relevant,
as many questions regarding their spin structure are still open \cite{Tomi}. The present work highlights the magnetic ordering in bulk and nanoparticles of MnCr$_2$O$_4$ chromite, focus has given to understand the origin of magnetic transition at $\sim$ 18 K in MnCr$_2$O$_4$. 
\section{Experimental Results} 
\subsection{Sample preparation and Characterization}
We have prepared MnCr$_2$O$_4$ nanoparticles by mechanical milling of the bulk material using Fritsch Planetary Mono Mill "Pulverisette 6". For bulk MnCr$_2$O$_4$, the stoichiometric mixture of MnO (99.999 \% purity) and Cr$_2$O$_3$ (99.997 \% purity) oxides were ground for 2 hours and pelletized. The pellet was sintered at 1000$^0$C for 12 hours and at 1200$^0$C for 24 hours with intermediate grinding. The sample was then cooled to room temperature at 2-3$^0$C/min. Before milling, the bulk sample was ground for two hours. X ray diffraction (XRD) spectra of the 
bulk sample was recorded using Philips PW1710 diffractometer with Cu K$_{\alpha}$ radiation. The XRD spectra of bulk sample matched with a 
typical spinel structure in cubic phase. The powdered MnCr$_2$O$_4$ (bulk) material was mechanical milled in a 80 ml agate vial with 10 mm agate 
balls with ball to sample mass ratio 12:1. The sintering and milling of bulk sample was carried out in Ar atmosphere to prevent the oxidation of Mn$^{2+}$. The milled samples are designated as mhX, where X denotes the number of milling hours. The XRD spectra of milled 
samples are also matched with standard cubic spinel structure. The XRD spectra, without any additional phase, showed that both bulk and nanoparticle samples are single phase chromites and having identical 
chemical composition. The absence of any impurity lines excludes the possibility of the formation of other alloyed compounds during mechanical milling. The lattice parameter of the samples were determined by standard 
full profile fitting method (Rietveld method) using FULLPROF Program. The calculations were performed assuming that the studied samples belong to Fd3m space group and ascribed to normal cubic spinel structure. The XRD peak lines were fitted with Lorentzian shape.
The lattice parameter of our bulk sample is 8.41($\pm$ 0.002)$\AA$, which is close enough to the reported value 8.42($\pm$ 0.02)$\AA$ \cite{Swad} and 8.437 $\AA$ \cite{Fava}. The systematic broadening of XRD lines with milling time corresponds to the decrease of particle size (Table I). The particle size was determined from the analysis of XRD 311 line using 
Debye-Scherrer equation. The effect of micro-strain, induced in the material during milling, in the determination of half-width has been minimized by matching the 311 line of the XRD spectra with Lorentzian 
shape. We have taken TEM data for three samples, i.e., mh60, mh48 and mh36. The TEM data (not shown in figure) give average particle size $\sim$ 11.5 nm, 12.7 nm and 14.4 nm in comparison with XRD data 11 nm, 12 nm and 13 nm for samples mh60, mh48 and mh36, respectively. This shows that the effect of mechanical strain is not significant in our determination of particle size using XRD data. It is found (Table I) 
that lattice parameter of the system is systematically increases with 
the decrease of particle size. This is reflected in the shift of the XRD peaks, as shown for 311 line in Fig. 1, to lower scattering angle (2$\theta$) with increasing milling time. We, further, examined the originality of the shift of XRD peak with decreasing particle size by annealing the mh60 sample at 600$^0$C (denoted as: mh60(t6)) and 900$^0$C (denoted as: mh60(t9)) for 12 hours under vacuum sealed condition. 
We have seen (inset of Fig. 1) 
that 311 peak of annealed sample tends toward the bulk sample. The decrease of lattice parameter, e.g. 8.474 $\AA$ (for mh60, particle size $\sim$ 11 nm), 8.442 $\AA$ (for mh60(t6), particle size $\sim$ 12.8 nm), 8.424 $\AA$ (for mh60(t9), particle size $\sim$ 17.9 nm) and 8.410 $\AA$ (for bulk sample), with increasing annealing temperature (i.e., increase of particle size) confirm the lattice expansion in 
our nanoparticles.
\subsection{DC magnetization}
\subsubsection{Temperature dependence of magnetization}
The temperature dependence of dc magnetization (Fig. 2) was measured under zero field cooled 
(ZFC) and field cooled (FC) modes, using SQUID (Quantum Design, USA) magnetometer.
In ZFC mode, sample was cooled from 300 K to 2 K in the 
absence of dc magnetic field, followed by the application of 
magnetic field at 2 K and magnetization data were recorded while increasing 
the temperature. In FC mode, the sample was cooled from 300 K in presence of  field. Cooling field and measurement fields are same. The zero field cooled magnetization (MZFC) at 100 Oe of bulk sample (Fig. 2a) shows a sharp increase below 45 K and remains almost temperature independent in the temperature range 40 K-18 K. Below 18 K, MZFC sharply decreases down to our measurement temperature 2 K. The magnetic behaviour of our bulk MnCr$_2$O$_4$ is consistent with the 
reported \cite{Tomi,Sum} data which showed paramagnetic to ferrimagnetic transition 
below T$_C$ $\approx$ 45 K and collinear spin structure in B sites becomes non-collinear (canted) below T$_{S}$ $\approx$ 18 K, due to the dominance of J$_{BB}$ interactions over J$_{AB}$ interactions \cite{Fava}. There is a small magnetic irreversibility (MFC $>$ MZFC) between field cooled magnetization (MFC) and MZFC below 40 K. The magnetic  
change observed in MZFC 
at T$_S$ $\approx$ 18 K also exists in MFC. It is noted that T$_S$ remains almost unchanged upto field $\sim$ 1 kOe, but get suppressed at 50 kOe data. This is consistent with a strong non-collinear ferrimagnetic order in bulk sample below 18 K, and such magnetic order is affected only at high magnetic field. \\
The M(T) (MZFC and MFC) data for nanoparticle (mh12, mh24, mh36, mh48 and mh60) samples
are shown in Fig. 2b-2f. The behaviour is different in comparison with the bulk. For example, magnetic irreversibility between MZFC and MFC starts well above of 45 K and the separation between 
MZFC and MFC increases on lowering the temperature. The other 
important observation is that plateau behaviour in the magnetization data 
(temperature range 40 K-18 K) of bulk sample slowly decreases and a 
maximum occurs below 40 K with decreasing the particle size (Fig. 2f). 
Note that there is no decrease of MFC below 18 K, instead the MFC in all nanoparticle samples show continuous increase down to 2 K. Fig. 2f (for mh60 sample) shows that 
the temperature T$_{irr}$, where magnetic irreversibility starts, decreases with increase of magnetic field. This behaviour of M(T) data along with large difference in MFC and MZFC shows blocking phenomenon of ferrimagnetic nanoparticles \cite{Chen}.\\
We, now, analyze the M(T) data to understand the particle size effect. Fig. 3a shows the temperature dependence of the inverse of dc magnetic susceptibility ($\chi$), using MZFC data at 100 Oe. It is interesting to note that $\chi^{-1}$ (T) of bulk and nanoparticle samples 
follow the typical functional form: $\chi^{-1}$ = T/C +1/$\chi_0$
-$\sigma$/(T-$\theta$), which consists of a Curie-Weiss (ferromagnetic) term and a 
Curie (paramagnetic) term and such equation has shown its application in 
ferrimagnetic spinel \cite{Yang}. This means strong ferrimagnetic order still exist 
in our nanoparticles. The extraction of the constants C and $\sigma$ form this equation is difficult because they are coupled in the paramagnetic regime. The asymptotic curie temperature ($\theta$), obtained by extrapolation as in Fig. 3a, is shown in the inset of Fig. 3. The closeness of T$_C$ 
and $\theta$ in bulk sample characterizes its long range 
ferrimagnetic order. 
On the other hand, $\theta$ values for nanoparticle sample (e.g. $\sim$ 60 K for particle size $\sim$ 19 nm, $\sim$ 70 K for particle size $\sim$ 16 nm) are much higher than the $\theta$ $\sim$ 45 K for bulk. The increasing magnetic disorder in nanoparticle is manifested by the large difference between $\theta$, and T$_C$ \cite{Will}, and 
broadening of the change in magnetization with temperature about T$_C$. According to
the core-shell model of ferrimagnetic nanoparticles \cite{Kodama}, the disorder is associated with increasing contribution of shell (surface) spins and long range ferrimagnetic order is associated with core spins.\\
From  the first order derivative of M(T) (both ZFC and FC) data (Fig. 4a), we find that there is a minimum below T$_C$ ($\sim$ 45 K) for bulk sample. This minimum for MZFC becomes broad for nanoparticle samples 
and occuring at higher temperature with respect to the bulk sample. The 
increasing broadness about the minimum is related to the disorder effects 
in nanoparticle samples. This is understood from the fact that the sample is magnetically more ordered in FC state than the ZFC state. The first order derivative of M(T) (Fig. 4a) shows a sharp peak near to 
T$_S$ $\sim$ 18 K for bulk sample, which is broadened for nanoparticle samples. 
This indicates that B site (non-collinear) spin configuration below 18 K has been   changed in nanoparticles. The difference between field cooled magnetization and zero field cooled magnetization ($\Delta$M= MFC-MZFC) vs T data (Fig. 4b) show 
that for bulk sample the $\Delta$M is very small in the temperature range 
40 K-18 K and increases below 18 K. For nanoparticle samples, $\Delta$M shows significant increase below 60 K with shape different from the bulk. The large magnitude of $\Delta$M over the whole temperature range, comparing the bulk sample, suggests  
increasing disorder effects in nanoparticle samples. On the other hand, shifting of the MZFC minimum of nanoparticle samples to higher temperature may indicate 
the increase of T$_C$ with decreasing the particle size. Due to the increasing flatness about 45 K in M(T) data and occurance of separation between MFC and MZFC well above 45 K, it is very difficult to estimate the exact T$_C$ values for nanoparticle samples. The spin wave theory can be applied, since ferrimagnetic order dominates in both bulk and nanoparticle samples. We have analyzed the MZFC data at 50 kOe, using the Bloch law: M(T)=M$_{0}$(1-$\beta$T$^{\alpha}$). Here, M$_0$ is the extrapolated value of saturation magnetization at 0 K, and $\beta$ and $\alpha$ are the constants. The change of M$_0$, Bloch exponent $\alpha$ and Bloch coefficient $\beta$ with particle size are shown in Table I, 
and in Fig. (5b), respectively. The saturation magnetic moment $\sim$ 30 emu/g (or magnetic moment $\sim$ 1.46 ($\pm$ 0.01)$\mu_B$ per formula unit) at 2 K of our bulk sample is consistent with the reported value 1.4-1.6 $\mu_B$ \cite{Wick}. It is interesting from the log-log plot of (M$_0$ -M(T)) vs T (Fig. 5a) that curves for all the samples intersect at 
the same temperature point $\sim$ 44 K. This observation is notable in 
the sense that this temperature point is close to the T$_C$ $\sim$ 45 K of the bulk sample and invariant with particle size down to $\sim$ 11 nm. The Bloch exponent $\alpha$ = 3.35$\pm$0.04 for bulk MnCr$_2$O$_4$ is large in comparison with $\alpha$ = 1.5 for a typical ferromagnet \cite{Lub}. 
The exponent is reducing and approaching to the typical value of 1.5 with the decrease of particle size. For example, $\alpha$ = 1.7$\pm$0.05 for sample of particle size $\sim$ 11 nm. The theoretical calculation, as well as some experimental results on fine particles and clusters  \cite{Xavier,Hendri} have shown that $\alpha$ becomes larger than 1.5, 
the value corresponding to bulk material. Hence, the 
particle size dependence of $\alpha$ is unconventional in MnCr$_2$O$_4$, where $\alpha$ decreases with decrease of particle size. 
The feature is that T$^{3/2}$ spin wave law alone can not be applied for bulk sample and there is a possibility of another origin contributing in M(T) behaviour. The extra magnetic contributions may introduce from a modulated magnetic order, arising from the 
non-collinear spin structure \cite{FePt} in bulk system or alternation 
of core-shell structure \cite{rnbprb2} in nanoparticles. The tendency of decreasing the exponent towards 1.5, a typical value for standard ferro or ferrimagnet, for smaller particles suggest that the extra magnetic contribution, introduced due to the non-collinear structure of B site spins, decreases with the decrease of particle size. 
\subsubsection{Field dependence of magnetization}
Fig. 6(a-c) show the M(H) data and Fig. 6(d-e) show the corresponding Arrot plot (M$^2$ vs H/M) for bulk (mh0), mh60 and mh48 samples, respectively. 
In bulk sample, an initial rapid increase of M to its spontaneous magnetization (M$_S$) value within 1 kOe is followed by the lack of saturation (steep increase) with field (H) upto 120 kOe, for all temperatures below T$_C$. For higher temperatures, {\it e.g.} T = 40 K 
and 50 K, the non-linear contribution of M(H) is increasing in our bulk sample. Similar M(H) behaviour in MnCr$_2$S$_4$ \cite{Tsur} has been attributed to the coexistence of ferrimagnetic state and quasi paramagnetic state, arising from the non-collinear spin structure. The non-collinear spin structure between B site Cr$^{3+}$ moments MnCr$_2$O$_4$ has been confirmed by neutron diffraction experiment \cite{Hasti}. The non-saturation of magnetization at higher field is attributed to 
paramagnetic type contributions, as shown in inset of Fig. 6a, arising 
from the B site non-collinear spin structure. The non-linear contribution in M(H) data is observed even at lower temperatures for nanoparticle samples, as shown in Fig. 6b for mh60. Fig. 6c suggest that a typical paramagnetic state (the linear increase of M(H)) for mh48 sample is observed only above 70 K. The gradual change in M(H) data near to T$_C$ 
and above makes it very difficult to determine the exact value of T$_C$.
In order to distinguish the long range order ferrimagnetic state from the paramagnetic state, we have analyzed the M(H) data using Arrot plot (M$^{2}$vs H/M). The Arrot plots are shown in Fig. 6d-f. We have calculated the magnitude of spontaneous magnetization (M$_S$) by extrapolating the data for H $\geq$ 10 kOe to the M$^2$ axis. Although M(H) data at 50 K (Fig. 6a) is non-linear for bulk sample, the Arrot plot (Fig. 6d) gives zero value of M$_S$ and confirms the paramagnetic state of bulk sample at 50 K. This is consistent with paramagnetic to ferrimagnetic transition temperature T$_C$ $\sim$ 45 K, observed from the M(T). Hence, non-linear increase of M(H) at 50 K (above T$_C$) can be attributed to the short range interactions amongst the spins or clusters of spins, as found in other system \cite{Mira}. The upward curvature in Arrot plot, even in the order magnetic state, is not very conventional. The gradual increase of upward curvature with increase of temperature above T$_C$ of mh48 
sample (Fig. 6f) suggests that the paramagnetic contribution or disorder effect causes such curvature in Arrot plot even in the ferrimagnetic state. The more curvature in nanoparticle samples in comparison with the bulk indicates more magnetic disorder for lower particle sample. The paramagnetic contribution from each isotherm at T$\leq$ T$_C$ were determined by fitting the data above 10 kOe either to linear equation (example 10 K data of bulk sample) or non-linear equation (example: 40 K data of bulk sample). We find that the value of the spontaneous magnetization (M$_S$), determined from the revised Arrot plot (not shown in figures), after the subtraction of the paramagnetic contribution is slightly higher than that obtained without subtracting the 
PM contribution.
Fig. 7 shows the temperature dependence of M$_S$ for bulk and nanoparticle samples. The immediate feature is that M$_S$ is decreasing for lowering 
particle size and M$_S$ is not zero above 45 K, as in bulk, for nanoparticles. By definition T$_C$, {i.e.}, paramagnetic to ferrimagnetic transition temperature, is the temperature below which non-zero value of M$_S$ exists in the sample. Therefore, Fig. 7 suggests the increase of T$_C$ in the temperature range 45 K-55 K with the decrease of particle size. To confirm the increase of T$_C$, we have performed ac susceptibility measurements for selected samples and the data are shown in Fig. 8. The sharp transitions (peak) both in $\chi^\prime$ (real part of ac susceptibility) and in $\chi^{\prime\prime}$ (imaginary part of ac susceptibility) near to T$_C$ and T$_S$ are observed (Fig. 8a, Fig. 8b) 
for bulk sample, but no such peak either at T$_C$ or T$_S$ in $\chi^\prime$ is seen for nanoparticle samples.  However, $\chi^{\prime\prime}$ shows a sharp peak which is independent of frequencies (Fig. 8d) and corresponds to the T$_C$ of the sample. The occurance of this $\chi^{\prime\prime}$ peak at higher temperature for smaller particle samples (Fig. 8b) confirms the increase of T$_C$ with decreasing particle size of MnCr$_2$O$_4$. A small (but clear) change in $\chi^{\prime\prime}$ near to T$_S$ $\sim$ 18 K  
for our smallest particle size sample indicates that signature of non-collinear spin transition still present there, but weak in magnitude. 
Although a broad maximum in $\chi^\prime$, resemble to superparamagnetic blocking, at about 35 K is noted for mh60 sample, the other observations i.e., almost no shift of $\chi^\prime$ maximum with frequencies in the range 37 Hz-637 Hz (Fig. 8c) and no such maximum in $\chi^{\prime\prime}$, indicates that such maximum in $\chi^\prime$ is simply due to the disorder effect of shell spins, but not due to a typical superparamagnetic behaviour. The exhibition of a sharp peak in $\chi^{\prime\prime}$ near to T$_C$ conclusively indicates that long range ferrimagnetic order still dominant in our smallest particle size (11 nm) sample, and coexists with the increasing disorder of shell spins. 
\section{Discussion}
It is interesting to see that MnCr$_2$O$_4$ nanoparticles show lattice expansion. 
A number of explanations for the lattice expansion in nano-materials, such as: change in oxygen coordination number with the cations \cite{Fava}, change of valence state of cations \cite{Tun}, crystallographic phase transformation \cite{Pabi},
and contribution of excess volume of grain boundary spins \cite{Ayub,Wang}
are available in the literature. However, there is no satisfactory explanation. According to Banerjee et al. \cite{Ayub} lattice 
expansion in mechanical milled samples may be related to the mechanical 
strain induced effect, rather than intrinsic properties of the sample. This may not be true, because lattice expansion has been observed in both mechanical milled nanoparticles as well in chemical route prepared nanoparticles. Details of the origin of lattice expansion in our nanoparticles are discussed below.\\ 
The XRD pattern of our nanoparticle samples are identical with the bulk samples. The absence of any additional lines with respect to standard cubic spinel structure indicated that there is no crystallographic phase transformation in our nanoparticle sample. 
The agreement of the magnetic parameters (magnetic moment and T$_C$) of our bulk sample with literature value \cite{Wick,Hasti,Sum} has confirmed that Mn ions are in divalent (Mn$^{2+}$: 3d$^{5}$) state. The mechanical milling in argon atmosphere is also not in favour of the formation of Mn ions with higher ionic (3+ or 4+) states. By comparing the outer shell spin configuration of Mn$^{2+}$ (3d$^5$, moment: 5 $\mu_B$), Mn$^{3+}$ (3d$^4$, moment:4 $\mu_B$) and  Mn$^{4+}$ (3d$^3$, moment:3 $\mu_B$), it is evident that if Mn$^{3+}$ or Mn$^{4+}$ exists, at all, in our nanoparticle samples, the decrease of lattice parameter was expected. The increase of lattice 
parameter in nanoparticle samples conclude that there is no change in valence state of Mn ions. Hence, the decrease of magnetic moment in nanoparticle samples may be consistent with core-shell model \cite{Kodama}. 
The core-shell model \cite{Kodama} suggests that disorder in shell (surface) spins decrease 
the net magnetization of ferrimagnetic nanoparticle and such contribution of shell spins 
increases on decreasing the size of particle. The questions naturally raised, whether shell spin disorder is responsible only for the decrease of magnetization or this disorder may be translated into the lattice dynamics which can show 'disorder induced' magnetic order \cite{Nature}. An extensive work is going on to understand the role of shell (surface) spin  disorder \cite{rnbprb2,rnbcozn,rnbprb3} in different kind of nanoparticle spinel.\\
We, now, consider 
the effect of shell spin disorder in MnCr$_2$O$_4$ nanoparticles. The microstructure of the shell may influence the lattice expansion in two ways, by increasing the free excess volume of the incoherent shell spins in the interface structure, and by lowering symmetry in oxygen coordination numbers with surface cations. F. Fava et al.\cite{Fava} have already shown  that increase of lattice volume in MnCr$_2$O$_4$ is related 
to the change in oxygen coordination number with the cations. Consequently, the lattice pressure on core spins may be reduced by the elastic coupling between (shell and core) spin lattices \cite{Ayub}. 
Many other factors such as breaking of long range crystallographic
coherent length and random orientations of shell spins may exhibit interatomic spacing (lattice parameter) which is different from bulk lattice. 
Experimental results show larger interatomic spacings (lattice parameter) in MnCr$_2$O$_4$ nanoparticle. Since the ratio of J$_{AB}$ and J$_{BB}$ superexchange interactions in chromites depend on both the bond angle and bond length (interatomic spacings) of B site spins (cations) \cite{Wick}, any change in the spin configuration either in shell or core must be reflected in the magnetic properties of nanoparticles. If the lattice expansion in MnCr$_2$O$_4$ nanoparticles is intrinsic, one would expect 
two important effects. First, direct cation-cation (antiferromagnetic) interactions will be diluted. Consequently, non-collinear spin structure between B site Cr$^{3+}$ cations of bulk MnCr$_2$O$_4$, revealed by the magnetic transition at $\sim$ 18 K \cite{Wick}, will be decreased in MnCr$_2$O$_4$ nanoparticles. Second, inter-sublattice (J$_{AB}$) super-exchange (ferrimagnetic in nature) interactions will dominate over B-B (J$_{BB}$) interactions (antiferromagnetic in nature) \cite{Tsur,Wick}. Consequently, increase of T$_C$ is expected, as T$_C$ is proportional to J$_{AB}$ in spinel. 
A strong relationship between the lattice parameter and T$_C$ of 
chromites has been found from the comparison of lattice parameters ($\sim$ 8.41 $\AA$, 10.18 $\AA$ and 10.23 $\AA$, respectively) vs T$_C$ ($\sim$ 45 K, 80 K and 84.5 K, respectively) for bulk 
MnCr$_2$O$_4$, MnCr$_2$S$_4$ and CdCr$_2$S$_4$ with identical lattice 
(cubic spinel) structure. 
The enhancement of T$_C$ in MnCr$_2$O$_4$ nanoparticles, therefore, suggests the increase of interatomic distance between B site (Cr-Cr) cations.\\
We, now, show that the enhancement of T$_C$ is not due to the 
site exchange of Mn$^{2+}$ and Cr$^{3+}$ amongst A and B sites in nanoparticles. Both Rh and Cr atoms are highly stabilized in B sites of cubic spinel structure due to their strong affinity for B site. The work on  mechanical milled CoRh$_2$O$_4$ nanoparticles \cite{rnbprb2} 
indicated that there is no migration of Rh atoms from B site to A site 
for particle size down to $\sim$ 16 nm. On the other hand,
very small amount of cations exchange amongst A and B sites in mechanical milled Zn$_{0.8}$Co$_{0.2}$Fe$_2$O$_4$ nanoparticles \cite{rnbcozn} have shown drastic enhancement of both magnetization and ferrimagnetic ordering temperature in comparison with their bulk counterpart. Similar type of enhancement in magnetization is also expected in MnCr$_2$O$_4$ nanoparticle, if mechanical milling really affect the site selection of 
cations (Mn$^{2+}$ and Cr$^{3+}$). In bulk MnCr$_2$O$_4$, A site 
is fully occupied by Mn$^{2+}$ ions and B site is fully occupied by Cr$^{3+}$ ions. 
Experimentally it has been found that 
magnetic moment of Mn$^{2+}$ and Cr$^{3+}$ ions are $\sim$
5$\mu_B$ and 3$\mu_B$, respectively. For the sake of argument 
we assume that some of the Mn$^{2+}$ ions migrate to B site in the exchange of equal number of Cr$^{3+}$ ions to the A site in 
MnCr$_2$O$_4$ nanoparticles. Considering the formula for total  magnetization of spinel oxide (M = M$_B$-M$_A$, where M$_B$ and M$_A$ are magnetization of B and A sites, respectively) and the fact that the number of B site magnetic ions is double in comparison with the number of 
A site magnetic ions, the drastic enhancement of total magnetization was expected in MnCr$_2$O$_4$ nanoparticles. However, experimental results 
show the decrease of magnetization with decreasing particle size. Hence, we exclude the possibility of the site exchange of Mn$^{2+}$ and Cr$^{3+}$ among A and B sites in MnCr$_2$O$_4$ nanoparticles. Therefore, Cr atoms even in nanoparticle show the great affinity to occupy the B site alone, as in bulk.\\
We,now, consider the particle size effect on the magnetic transition at $\sim$ 18 K. Neutron experiments \cite{Tsur,Hasti} confirmed that collinear to non-collinear (canted) spin transition is the origin of magnetic transition at $\sim$ 18 K in bulk MnCr$_2$O$_4$. Further increase of surface spin canting should result in more prominent magnetic transition at $\sim$ 18 K, and decrease of ferrimagnetic ordering temperature (T$_C$)  for nanoparticles. Our experiment on MnCr$_2$O$_4$ nanoparticles show neither strong magnetic transition at 18 K nor decrease of T$_C$ for smaller particles. On the other hand, experiment confirmed the enhancement of T$_C$ in MnCr$_2$O$_4$ nanoparticles. Hence, weakening of the magnetic transition at 18 K in nanoparticles is not due to the effect of increasing surface spin canting. This is 
an effect of intrinsic change in non-collinear to collinear spin structure in B 
sites of MnCr$_2$O$_4$ and directly correlated with lattice expansion of the system \cite{Fuk}. \\
The other observations of nanoparticles, i.e. (i) large magnetic irreversibility between MFC and MZFC, (iii) appearance of a maximum in MZFC below 40 K, are resemble to superparamagnetic like blocking of nanoparticles. But, ac susceptibility measurement shows no typical superparamagnetic behaviour in our nanoparticles. Such behaviour, in fact,  arises due to the increasing disorder of surface spins. The intersection of M(T) data, following Bloch law, near to 44 K (close to T$_C$ $\sim$ 45 K of bulk sample) for all samples suggests that strong ferrimagnetic order of core spins (bulk) is almost retained.  
We, therefore, suggest that surface spin disorder in nanoparticles is translated into the lattice dynamics to cause geometrical frustration effect \cite{Nature} and results in lattice expansion. 
\section{Conclusions}
Bulk MnCr$_2$O$_4$ is a ferrimagnet with paramagnetic to collinear ferrimagnetic state at T$_C$ $\approx$ 45 K and collinear ferrimagnetic to non-collinear ferrimagnetic state below 18 K. Experimental results suggest that non-collinear ferrimagnetic state in MnCr$_2$O$_4$ occurs due to direct interactions between B site Cr$^{3+}$ ions. The B site direct interactions, represented by magnetic transition at 18 K, decreases
in MnCr$_2$O$_4$ nanoparticles. We attribute the decrease of B site direct interactions to the lattice expansion in MnCr$_2$O$_4$ nanoparticles, essentially confined in shell, which results in the change from non-collinear to collinear structure of B site spins. In this sense, our experimental work provide substantial evidence to, further, confirm the proposal made by Wickham and Goodenough that direct (antiferromagnetic) interactions between B site cations are possible in chromites and causes collinear to non-collinear ferrimagnetic transition at about 18 K. 
Reduction of magnetic moment, large magnetic irreversibility between MZFC 
and MFC, and appearance of superparamagnetic like blocking are some of the notable 
disorder effects in nanoparticles. Exceptional large Bloch exponent 
$\alpha$ = 3.35 for bulk MnCr$_2$O$_4$ and decrease of $\alpha$ with 
decreasing particle size, identify unconventional magnetic 
features in MnCr$_2$O$_4$. 

\vskip 0.3 cm
\noindent Acknowledgment:
We thank Dr. Chandan Mazumdar for useful discussions and 
SQUID measurements.

\begin{table*}
\caption{\label{tab:table1} The parameters for milled sample mhX, where X represents milling time in hours on bulk sample. particle 
in size (D) and lattice parameter a($\AA$) were obtained from XRD spectra. Saturation magnetization at 0 K (M$_0$) was obtained from the M(T) data at 
50 kOe. T$_C$ was determined from first order derivative of M(T), M$_S$ (T) and ac susceptibility data.}
\begin{ruledtabular}
\begin{tabular}{ccccc}
Sample & D (nm) & a($\pm$0.002 $\AA$) & T$_{C}$ (K) & M$_0$(emu/g) \\\hline
mh0 & 150 & 8.410 & 44$\pm$1 & 30.54\\
mh12 & 19 & 8.412 & 46$\pm$1 & 25.91\\
mh24 & 16 & 8.425 & 47$\pm$1 & 25.24\\
mh36 & 13 & 8.440 & 49$\pm$1 & 22.81\\
mh48 & 12 & 8.448 & 51$\pm$1 & 22.35\\
mh60 & 11 & 8.474 & 52$\pm$1 & 21.78\\ 
\end{tabular}
\end{ruledtabular}
\end{table*}

\end{document}